# Chiral optics without chiral matter

**Jingxuan Wei**[1]

[1] State Key Laboratory of Electronic Thin Films and Integrated Devices, School of Optoelectronic Science and Engineering, University of Electronic Science and Technology of China, Chengdu 611731, China

## Key message

The effects of chiral light can manifest in matter as differential scalar, vector, or pseudovector responses, even in the absence of geometric chirality.

## Abstract

Strong chiral light-matter interactions are conventionally believed to require structurally chiral materials. However, emerging evidence challenges this assumption, prompting a fundamental question: do chiral matters really matter in circularly polarized light (CPL) responses? While chiral structures favor circular dichroism generation in isotropic media, non-chiral systems with anisotropy can generate circular dichroism, and CPL can drive directional charge transport and spin-selective processes without structural chirality. By unifying these phenomena through fundamental light-matter interaction principles, recognizing that light carries both chiral and achiral components while matter exhibits diverse degrees of freedom such as charge, moment and angular momentum, we reveal a broader landscape of chiral optics. This perspective reconciles seemingly disparate findings and opens pathways to advanced technologies including high-performance CPL devices, ultrasensitive chiral sensors, and spin-selective photocatalysts, fundamentally expanding chiral optics beyond traditional structural constraints.

## Representative references

## To make a long story short

The interaction between chiral light (circularly polarized light, CPL) and matter represents a fundamental process with broad implications across numerous scientific disciplines. Conventionally, it has been held that strong chiral light-matter responses necessitate the use of structurally chiral materials. Yet, emerging evidence increasingly challenges this perspective. We therefore revisit a central question: **must matter be chiral to matter**?

In brief: while chiral structures are indeed essential for generating circular dichroism (CD) signals in isotropic media containing randomly oriented molecules, **chirality is not a prerequisite across numerous other scenarios**. CD, a scalar absorption asymmetry, can arise in achiral systems that exhibit anisotropy and orientational order, for example, aligned molecular assemblies or anisotropic crystals. Moreover, CPL can drive vectorial responses (such as directional charge transport) and pseudo-vectorial responses (such as spin- or valley-selective population) without any structural chirality in the medium.

In view of these considerations, we introduce a unified theoretical framework grounded in the fundamental principles of light-matter interaction. By decoupling chiral optical responses from structural chirality, this expanded framework enables a broad class of technologies, including high-performance CPL detectors and emitters, ultrasensitive chiral sensors, and spin-selective photocatalysts.

## Main text

Chirality describes a geometric property where an object or system cannot be superimposed onto its mirror image through any combination of translations or rotations. This fundamental concept permeates multiple scales from microscopic particles to macroscopic objects and plays a pivotal role across various disciplines including physics, chemistry, and biology. In optics, circularly polarized light (CPL) exhibits pronounced chiral characteristics, with its electric and magnetic field vectors tracing a helical trajectory along the propagation axis, categorized into left-handed and right-handed circular polarization (LCP and RCP) states based on the rotation direction. This unique light polarization finds significant applications in cutting-edge fields such as chiral molecular detection, quantum communication, display technologies, remote sensing, astronomical observation, and photocatalysis.

Conventional wisdom holds that distinguishing between LCP and RCP light requires the use of chiral objects or measurement systems[1]. This is particularly true in isotropic media containing randomly oriented molecules, where differential absorption of LCP versus RCP light emerges exclusively when the molecules possess chiral structures[2]. This phenomenon, known as circular dichroism (CD), is linked to the optical chirality $\mathrm{Im}(\boldsymbol{E}^{*}\cdot\boldsymbol{H})$, a quantity that highlights the coupling of electric and magnetic fields[3].

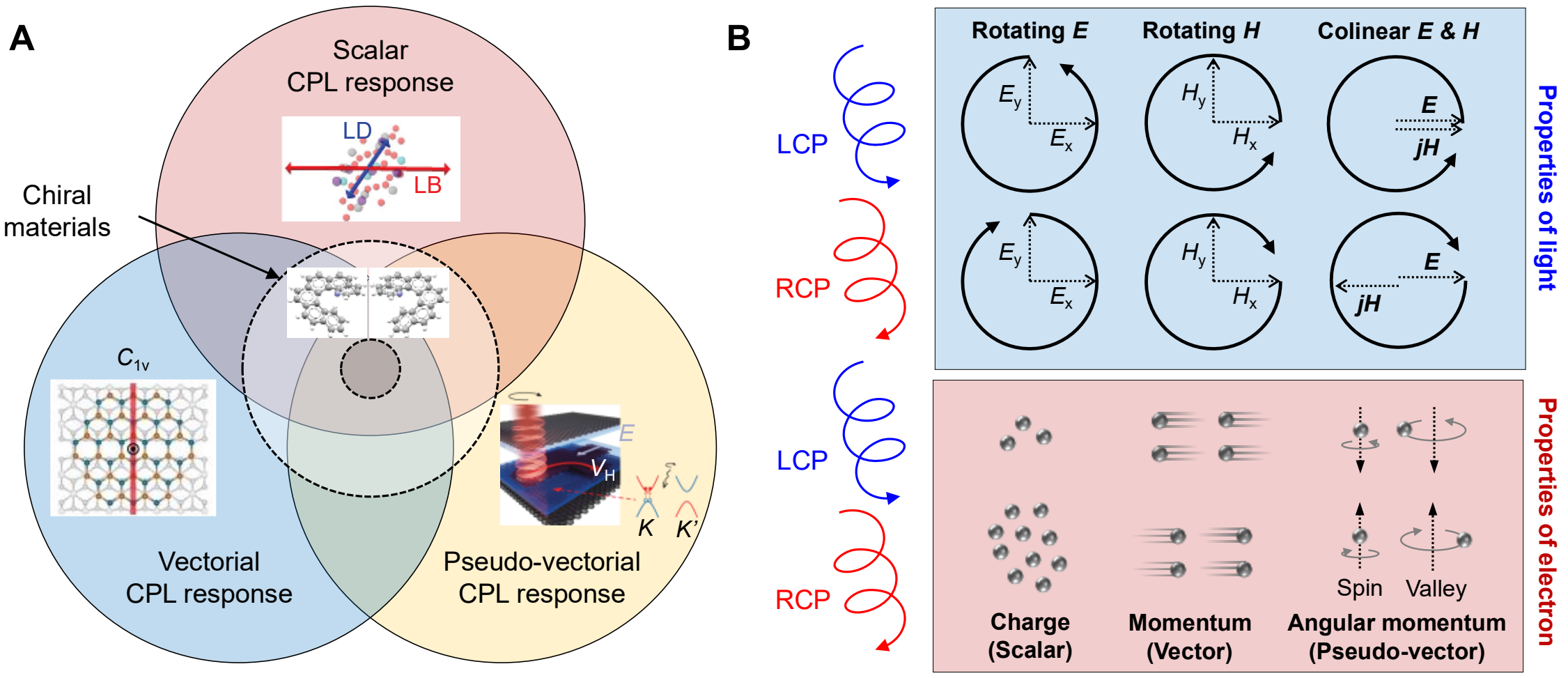

**Figure | Circularly polarized light (CPL)–matter interactions across chiral and achiral platforms.** (**A**) Representative classification of reported CPL responses into scalar, vectorial, and pseudo-vectorial types, illustrating that geometric chirality is not required for all cases. (**B**) Light-matter interaction view highlighting how helicity-dependent fields can yield differential responses in charge density (scalar), momentum (vector), and angular momentum (pseudo-vector), as documented in both achiral and chiral systems.

However, this conventional view does not hold in other cases (see the figure, **panel A**). A striking example is the recent observation of CD in $Li_2Co_3(SeO_3)_4$, a centrosymmetric (and thus achiral) material[4]. Here, optical axis tilting breaks in-plane symmetry via electric dipole-mediated interactions, generating CPL-selective absorption through the cooperative effects of linear dichroism and linear birefringence (LDLB). Indeed, theoretical work has long demonstrated that CD in non-chiral systems originates from electric-electric coupling, distinct from the electric-magnetic coupling underlying the effect in chiral molecules[2]. This phenomenon occurs in systems with preferred orientations, such as crystals or other anisotropic media, where the symmetry conditions fundamentally differ from those in isotropic chiral systems.

Beyond the scalar effect of differential absorption, CPL can also induce vectorial effects (e.g., directional charge motion) without a strict requirement for chirality[5]. This vectorial CPL-response imposes fundamentally different symmetry constraints than conventional CD: while mirror symmetry may be preserved, all in-plane $C_2$ and higher-order rotational symmetries must be broken, where geometric chirality becomes irrelevant. Such vectorial responses emerge through distinct optoelectronic mechanisms including the bulk photovoltaic effect, circular photogalvanic effect, and photon drag effect. These phenomena have now been observed across diverse quantum materials systems, from topological semimetals and ferroelectrics to moiré superlattices[6].

Similarly, CPL can generate a pseudo-vector asymmetry, exemplified by the imbalanced population of electron spins or valleys, which constitutes a third class of chirality-independent response[7–9]. Unlike scalar or vector quantities, these pseudo-vector characteristics exhibit distinct transformation properties under symmetry operations. Hexagonal two-dimensional materials with $C_{3v}$ symmetry ($MoS_2$, Graphene) exemplifies this principle[8]: while mirror symmetry eliminates scalar responses and rotational symmetry cancels vector components, pseudovector phenomena persist under the combined threefold rotational and mirror symmetries. The circular polarization-dependent pseudo-vector properties of spin and valley degrees of freedom can be converted into measurable electrical signals through the inverse spin Hall effect and valley Hall effect[8]. Under longitudinal bias, this manifests as transverse separation of charge carriers with distinct spin or valley indices, generating spin- or valley-polarized currents. Importantly, the boundary between these response types is strictly governed by the underlying crystal symmetry. While the valley-selective optical transition manifests as a pseudo-vectorial response requiring an external bias in $C_{3v}$ structures, further symmetry reduction (e.g., down to $C_1$ symmetry) can directly convert this asymmetric valley population into a zero-bias vectorial photocurrent.

The relationship between CPL responses and chirality can be elucidated from the standpoint of fundamental light-matter interactions. Building on the established classification of responses into scalar, vectorial, and pseudo-vectorial differences (see the figure, **panel B**), this framework systematically links the electromagnetic features of chiral light, such as rotating electric and magnetic fields and colinear electric-magnetic components, to the resulting electronic responses in materials, encoded in charge, momentum, spin, and valley degrees of freedom. In doing so, it consolidates prior observations into a coherent picture and provides a common language for interpreting past studies.

To elucidate the electromagnetic basis of CPL responses, we summarize the standard properties of CPL. For a monochromatic CPL propagating along the *z*-axis, the complete electromagnetic state can be characterized by four complex field components: ($E_x$, $E_y$, $H_x$, $H_y$), where $E$ and $H$ represent the electric and magnetic fields, respectively. Adopting the phase convention where $E_x$ has zero relative phase ($E_x = 1 + 0j$), the field components are ($1, j, j/\eta, 1/\eta$) for LCP and ($1, -j, -j/\eta, 1/\eta$) for RCP, where $\eta$ is the wave impedance of the medium. All components between LCP and RCP have identical amplitudes, and the only distinction between LCP and RCP lies in their relative phases. Specifically, CPL-response emerges exclusively through the interference of at least two field components[2]. If we limit the number of components to two, there will be three cases, which correspond to three properties of CPL: a rotating electric field[9], a rotating magnetic field, and a collinear electric-magnetic field[3]. Notably, while the rotating electric field ($E$ is a true vector) and rotating magnetic field ($H$ is a pseudovector) are each achiral, their colinear combination generates optical chirality. This reveals two distinct origins for CPL responses: chiral field combinations ($E$ plus $H$) or achiral field components ($E$ or $H$ alone). Hence, material chirality is not a prerequisite for CPL responses, provided the relevant symmetries permit them.

On the material side, we cast the viewpoint on how CPL information is encoded in electronic degrees of freedom beyond charge. The classical CD absorption microscopically arises from distinct populations of excited electrons governed by charge distributions. Importantly, electrons exhibit additional quantum degrees of freedom beyond charge, including momentum, spin angular momentum, and valley pseudospin. These enable more sophisticated encoding of helicity information through vectorial responses[5] (e.g., directional photocurrents via momentum-selective excitation) or pseudo-vectorial responses[7,8] (e.g., spin or valley polarization through selective population). As demonstrated in previous examples, these mechanisms allow CPL-matter interactions to extend far beyond traditional chiral systems.

The growing body of work on CPL responses of achiral systems has opened new avenues for CPL detection beyond traditional chiral-material approaches. Conventional chiral-material-based schemes face fundamental constraints: (1) Natural chiral molecule synthesis requires complex asymmetric catalysis with inadequate purity control, while artificial chiral nanostructures (e.g., helical nanowires, chiral photonic crystals) necessitate high-precision fabrication that is both cost-prohibitive and incompatible with standard semiconductor processes; (2) Molecular CD signals remain intrinsically weak due to polarizability limits, even when enhanced through supramolecular assembly; (3) CD activity is typically confined to UV-Vis ranges and degrades under extreme conditions (e.g., high temperature, radiation), restricting broadband utility. Conversely, the generalized CPL response framework, which encompasses scalar, vector, pseudovector, and matrix manifestations, enables versatile detector design principles. Vectorial CPL responses permit detectors with photocurrent linearly proportional to circular polarization degree, achieving intensity and linear-polarization independence[10]. Similarly, pseudo-vectorial effects enable direct spin/valley-polarized optoelectronic conversion, providing critical components for spintronic and valleytronic applications[8].

Similar considerations apply to CPL sources. Traditional emitters that rely on chiral molecules or intricate chiral nanostructures often exhibit limited polarization purity, narrow emission bands, and poor process compatibility. Studies of CPL generation in achiral media indicate alternative strategies based on engineered light-matter interactions. For example, controlled orientation in nominally achiral materials can yield CPL via coupling between linear dichroism and birefringence[11]; vectorial responses can support bias-switchable CPL emission through orbital-momentum locking[12]; pseudo-vectorial responses in transition-metal dichalcogenides offer compact, high-purity sources compatible with mainstream semiconductor processing[13].

These same mechanisms have also been leveraged in chiral sensing and photocatalysis. In molecular sensing, nanostructures designed to support strong interactions can create superchiral near fields that enhance chiral asymmetries by orders of magnitude, enabling interactions with randomly oriented molecules and improving detection limits beyond those of conventional CD spectroscopy[14]. In photocatalysis, vectorial responses manifest as directional photocurrents that establish internal electric fields and can improve charge separation and catalytic efficiency, while CPL-induced spin polarization offers additional control over spin-dependent reaction pathways, which is an emerging direction highlighted in recent catalytic studies[15].

Together, these observations clarify how achiral platforms can enable robust CPL detection, emission, sensing, and catalysis, thereby complementing and, in many cases, circumventing the traditional reliance on structural chirality.